\documentclass[aps,pdftex,superscriptaddress,prfluids,notitlepage,nofootinbib]{revtex4-2}
\usepackage{multirow,enumerate,epsfig,graphics,amssymb,amsmath,subeqnarray,mathrsfs,epstopdf}
\usepackage{epsfig,graphics,amssymb,amsmath,subeqnarray,graphicx,amsthm,subfigure,mathrsfs}
\usepackage[colorlinks=true, citecolor=blue, linkcolor=black]{hyperref}
\usepackage{tabularx}
\usepackage{upgreek}
\usepackage[utf8]{inputenc}
\usepackage[english]{babel}
\usepackage[]{graphicx,tabularx}
\usepackage{epstopdf}
\usepackage{lipsum}

\newcommand{\xb}{\boldsymbol{x}}
\newcommand{\ub}{\bar{\boldsymbol{u}}}
\newcommand{\Ub}{\boldsymbol{U}}
\newcommand{\fb}{\boldsymbol{f}}
\newcommand{\gb}{\boldsymbol{g}}

\newcommand{\Ca}{\mathrm{Ca}}
\newcommand{\bnabla}{\boldsymbol{\nabla}}
\newcommand{\Rey}{\mathrm{Re}}
\newcommand{\We}{\mathrm{We}}
\newcommand{\pard}[3][]{\ensuremath{\frac{\partial^{#1} {#2}}{\partial {#3}^{#1}}}}
\newcommand{\pardn}[3]{\frac{\partial^{#1} {#2}}{\partial {#3}^{#1}}}

\newcommand{\pderiv}[3][]{\ensuremath{\frac{\partial^{#1} {#2}}{\partial {#3}^{#1}}}}

\begin{document}

\title{Weak-inertial effects on destabilized receding contact lines}% Force line breaks with \\
%\thanks{A footnote to the article title}   %

\author{Akhil Varma} 
\email{varma@pks.mpg.de, akhil@varma.net}
\affiliation{Max Planck Institute for the Physics of Complex Systems, N\"othnitzer Str. 38, Dresden 01187, Germany}

\date{\today}% It is always \today, today,

\begin{abstract}
It is known that beyond a critical speed, the straight contact line of a partially-wetting liquid destabilizes into a corner. In one of the earliest theoretical works exploring this phenomenon, [\citeauthor{Limat04}, \textit{Europhys. Lett.} \textbf{65}(3), 2004] elicited a self-similar conical structure of the interface in the viscous regime. However, noting that inertia is not expected to be negligible at contact line speeds close to, and beyond the critical value for many common liquids, we provide the leading-order inertial correction to their solution. In particular, we find the self-similar corrections to the interface shape as well as the flow-field, and also determine their scaling with the capillary number. We find that inertia invariably modifies the interface into a cusp-like shape with an increased film thickness. Furthermore, when incorporating contact line dynamics into the model, resulting in a narrowing of the corner as the contact line speed increases, we still observe an overall increase in the inertial contribution with speed despite the increased confinement. 
\end{abstract}

\maketitle

\section{Introduction}
\label{sec1}

There is something intriguing about raindrops sliding down a window pane that fascinates us all on a rainy day. A keen observer might notice the drops speeding up as they slide down under gravity, forming long tails or rivulets that eventually break into smaller drops. This is a classic example of contact line destabilization. The capillary number, which is the dimensionless velocity of the contact line formed by the liquid interface and the solid surface, is the key player in this phenomenon; It is defined as $\Ca = \eta V/\gamma$, where $\eta$ is the viscosity of the liquid, $\gamma$ is its surface tension and $V$ is the velocity of the moving contact line. It is beyond a critical value $\Ca_\mathrm{cr}$, that a straight contact line destabilizes and starts to entrain air (at the advancing front) or form cornered tails (at the receding end). For many common liquids, the critical value at the receding contact line is $\Ca_\mathrm{cr} \sim \mathcal{O}(10^{-3})$ .

The instability of receding contact lines of partially wetting liquids were addressed as early as the works of \citet{Blake79,Petrov85}. They observed that beyond a critical capillary number, a straight receding contact line transforms into a wedge or corner having some semi-opening angle $\phi$. The slanted orientation of the contact line keeps its normal component of velocity below the threshold for wetting transition. Furthermore, they observed that exceeding the critical capillary number did not alter the normal velocity of the (destabilized) contact line. Instead, it consistently maintained the critical value by adjusting its opening angle accordingly \cite{Petrov85}. This is given by the simple phenomenological relation, $\sin \phi \propto 1/\Ca$. %In this manner, they identified an upper limit on the speed at which dewetting occurs.

Since this work, many experimental and theoretical studies followed which aimed at understanding the instability \cite{Podgorski01,Delon07,Snoeijer13} as well as the detailed shape of the destabilized contact line \cite{LeGrand05,Puthenveettil13,He_Nagel19}. However, the mechanism of this instability is not yet fully understood, and is still an active area of research \cite{Vandre13,Keeler22}. A comprehensive understanding of the contact line dynamics is indispensable for many engineering applications ranging from precision thin film coatings, %used in optics and automotive industries, processes such as 
immersion lithography, %in electronics, and controlled deposition in 
nano-fabrication and ink-jet printing \cite{Lohse22,Winkels11} %Naturally, this understanding also aids in
to the design of liquid-repellent and anti-fogging surfaces \cite{Bonn09}. They are also crucial for microfluidics and lab-on-chip technologies \cite{Stone04}.

For a drop of viscous liquid moving down an incline, the instability is observed to occur first at the receding (rear) end, giving rise to the classical "tear drop" appearance. At velocities much higher than this critical value, the sharp corner transforms into a cusp, eventually breaking into drops at even higher velocities through a pearling instability \cite{Podgorski01,LeGrand05,Itai02}. The flow close to the corner is viscous-dominated and hence can be approximated as a Stokes flow. This realization led \citet{Limat04}, and later \citet{Snoeijer05} to determine the interface profile and the flow field near the receding contact line. A thin-film approximation, assuming a slowly varying interface profile near the corner was assumed for their analysis. The theory matches well with the flow field observed in experiments with silicone drops \cite{Snoeijer05}. 
However, for fast-moving contact lines, inertial effects can be significant \cite{Stoev99,Puthenveettil13}. For example, consider water and silicone oil $(100\;\mathrm{cP})$ drops having a critical capillary number $\Ca_\mathrm{cr} \approx 4\times 10^{-3}$ and liquid mercury, having $\Ca_\mathrm{cr} \approx 1.5\times 10^{-3}$. The drop velocities in experiments of sliding drops are typically $10-100\;\rm{cm.s^{-1}}$ for water and mercury, and $0.1-1 \;\rm{cm.s^{-1}}$ for silicone oil \cite{LeGrand05,Puthenveettil13}. For these velocities the Reynolds number, which compares the inertial with the viscous effects, at distances of $l=10-100\rm{\mu m}$ from the corner turn out to be $\Rey_l \sim 0.01-0.1$ for silicone oil, $\sim 1-10$ for water and $\sim 10-100$ for mercury. The low Reynolds number for silicone oil probably explains why the model with Stokes flow approximation agrees well with experiments. However, the large Reynolds number in the case of water and mercury drops indicate that inertial effects are dominant, even amounting to orders of $10^3$ at the scale of the drops \cite{Puthenveettil13}. Thus, it seems that inertia cannot be excluded in the models of moving contact lines of these liquids. Prior theoretical works in this area have largely focused on the role of inertia near straight contact lines \cite{Cox98,Tak_shing_Chan_thesis,Varma21} or fully wetted films \cite{deRyck98,Koulago95}, but studies on cornered contact lines are scarce. For dewetting corners, \citet{Kim_Westerweel15} suggested using an effective dynamic pressure in the classical lubrication theory to mimic inertia. However, this results in a solution that is qualitatively indifferent from the non-inertial case. Here, we will use a regular perturbation approach to systematically determine the inertial contribution and establish the correct scaling associated with it.

In this theoretical work, we shall focus on determining the inertial effects solely on \emph{receding} destabilized contact lines. The mechanism of the instability is in itself quite complex, and is beyond the scope of this work. Thus, for our analysis, we shall presuppose that the contact line is destabilized, and forms a corner on the solid surface, as observed in experiments. Our objective then is to determine the shape of the interface and flow-field within. To this end, we closely follow the analyses of \citet{Limat04}, and \citet{Snoeijer05} and improve on their solution by introducing weak-inertial corrections in \S \ref{sec2}. In \S \ref{sec3} we solve the boundary value problem for the interface profile, which gives us the leading-order effect of inertia near the contact line. The self-similar form of the inertial correction is obtained here. %The flow-field near the cornered rear is of particular interest as it introduces flow in the third dimension due to lateral confinement from the contact line. 
Taking water and mercury as example liquids, the inertial corrections to their interface shape and the flow field are discussed in \S \ref{sec4}. Finally, a summary of our key findings and relevant discussions are offered in \S \ref{sec5}.

\section{Problem formulation}\label{sec2}

\begin{figure}[htbp]
    \centering
    \includegraphics[width=0.45\columnwidth]{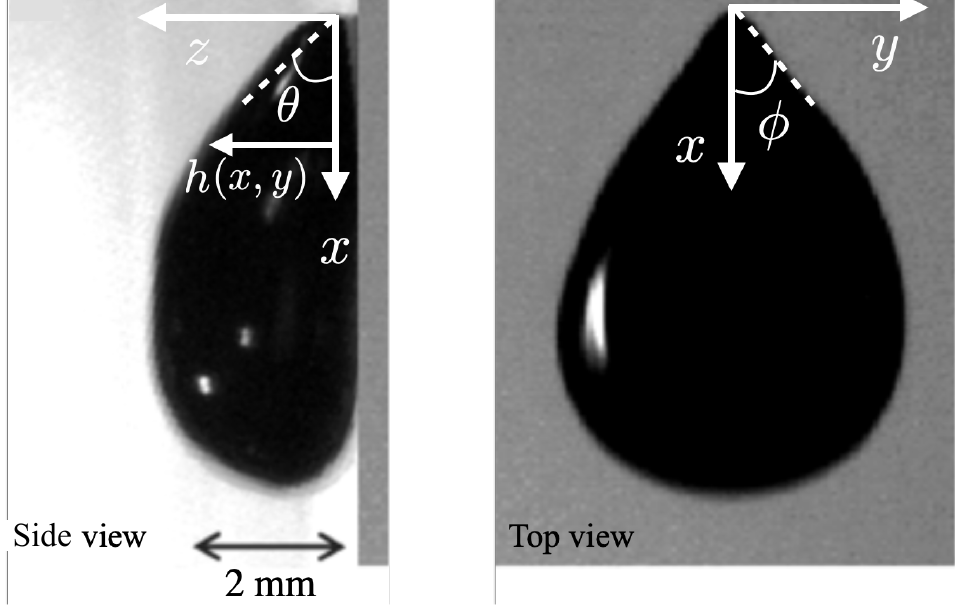}
    \caption{Coordinate system used for the analysis of destabilized receding contact line. Shown here are the side and top view snapshots from the experiments of \citet{Puthenveettil13} of a mercury drop sliding along the $x$ direction on a glass surface. The liquid interface is at $z=h(x,y)$ and gravity acts vertically down . The semi-opening angle of the contact line and the local side angle of the interface are denoted by $\phi$ and $\theta$ respectively. Dark grey is the solid surface (Adapted with permission from \cite{Puthenveettil13}.)}
    \label{fig:schematic}
\end{figure}

Consider the steady motion of the contact line of a partially-wetting liquid over a solid surface (at $z=0$ plane) with a velocity $V$. As an illustrative example, we take the case of a drop sliding along the $x$-axis as shown in Fig.\ref{fig:schematic} . 
Let the liquid have a density $\rho$ and viscosity $\eta$. The shape of the liquid interface is described by its height from the solid surface, $h(x,y)$. Close to the receding contact line, we assume a gradual variation of the interface height. We also assume the characteristic height of the interface, $a$, to be much smaller than the other dimensions which have a characteristic length $l$. This allows for a separation of scales between the planar ($x,y$) and the normal ($z$) directions, which is done by introducing the small parameter, $a/l = \delta \ll 1$. 
We first rescale all the physical quantities with their respective characteristic values to obtain the corresponding dimensionless quantities (denoted by $\bar{\bullet}$). For example, the dimensionless coordinates $(\bar{x},\bar{y},\bar{z}) = (x/l,y/l,z/a)$, velocity field $(\bar{u}_x,\bar{u}_y,\bar{u}_z) =(u_x/V, u_y/V, u_z/(\delta V))$ and pressure field $\bar{p}=p\delta^2 l/(\eta V)$. Note that we have rescaled pressure using the lubrication pressure, rather than inertial, as we are interested in predominantly viscous flows.  

\subsection{Momentum conservation}
\label{sec2a}

The steady-state momentum conservation within the liquid is given by the Navier-Stokes equation, written in planar $(x,y)$ and normal ($z$) directions respectively as,

\begin{align}
%\pderiv[2]{\ub}{z} - \bnabla p = \Rey\; (\ub \cdot \bnabla) \ub,
\delta^2 \bar{\Delta} \ub + \pderiv[2]{\ub}{\bar{z}} - \bar{\bnabla} \bar{p} & = \delta^2 \Rey_l \left(\ub \cdot \bar{\bnabla} \ub + \bar{u}_z \pderiv{\ub}{\bar{z}}\right), \label{eq:1} \\
\delta^2 \pderiv[2]{\bar{u}_z}{\bar{z}} - \pderiv{\bar{p}}{\bar{z}} & = \delta^4 \Rey_l \left(\ub \cdot \bar{\bnabla}\bar{u}_z + \bar{u}_z \pderiv{\bar{u}_z}{\bar{z}}\right). \label{eq:2}
\end{align}
Here, $\ub=(\bar{u}_x,\bar{u}_y)$, $\bar{\bnabla}=(\partial/\partial \bar{x}, \partial/\partial \bar{y})$ is the planar gradient, and $\bar{\Delta} = \partial^2 /\partial \bar{x}^2 + \partial^2 /\partial \bar{y}^2$ is the planar Laplacian. $\Rey_l = l V\rho/\eta$ is the Reynolds number based on the larger length scale, $l$. Classical lubrication theory ignores the inertial term by assuming small values of $\Rey_l \lesssim 1$. In contrast, here we consider the regime where $ \Rey_l \gg 1$, so as to retain the contribution of inertia. However, when the Reynolds number is sufficiently large, typically $\Rey_l \gtrsim \delta^{-2}$, boundary layer separation starts coming into the picture \cite{Schlichting00}. To avoid this complication, we limit our analysis to the moderate inertial regime defined by $1 \ll \Rey_l \ll \delta^{-2}$, i.e. $\Rey_l \sim \mathcal{O}(\delta^{-1})$. So now, by neglecting terms of $\mathcal{O}(\delta^2)$ and smaller, \eqref{eq:1} and \eqref{eq:2} reduce to
\begin{align}
     \pderiv[2]{\ub}{\bar{z}} - \bar{\bnabla} \bar{p} & = \epsilon\; (\ub \cdot \bar{\bnabla}) \ub, \qquad \mathrm{and}
\label{eq:NS} \\
\pderiv{\bar{p}}{\bar{z}} & = 0.
\label{eq:PZ}
\end{align}
%where we have used $\epsilon$ to denote $\epsilon \Rey \sim \mathcal{O}(\epsilon)$, a small parameter. Nevertheless,
where $\epsilon = \delta^2\;\Rey_l \sim \mathcal{O}(\delta)$ is a small parameter. Comparison with full numerical simulations of channel flow have shown the ``inertial thin-film equations" in \eqref{eq:NS} and \eqref{eq:PZ} to be accurate even for much higher values of $\epsilon (\sim 1-10)\;$ \cite{Jacono05}. To obtain the velocity field $\ub$, one has to integrate the expression \eqref{eq:NS} between $\bar{z}=0$ and $\bar{z}=\bar{h}(\bar{x},\bar{y})$, with the following boundary conditions in the laboratory frame-of-reference: (i) no-slip at the solid surface, $\ub\vert_{\bar{z}=0}=\boldsymbol{0}$, and (ii) free-shear at the liquid interface, $\partial \ub/\partial z \vert_{\bar{z}=\bar{h}} =\boldsymbol{0}$. Furthermore, it follows from these boundary conditions that $\bar{u}_z=0$.

To obtain an analytical solution in the limit of of small, but non-negligible values of $\epsilon$, we resort to a regular perturbation expansion of the form,
\begin{align}
    \ub&=\ub_0+\epsilon\ub_1 + \mathcal{O}(\epsilon^2), \quad \bar{h}=\bar{h}_0+\epsilon \bar{h}_1+ \mathcal{O}(\epsilon^2)\quad \mathrm{and}, \nonumber \\ \bar{p}&=\bar{p}_0+\epsilon \bar{p}_1+ \mathcal{O}(\epsilon^2).
    \label{eq:expn}
\end{align}
We truncate the expansion at $\mathcal{O}(\epsilon)$, the leading-order inertial term; The subscript $0$ denotes $\mathcal{O}(1)$ terms (Stokes flow) and $1$ denotes $\mathcal{O}(\epsilon)$ terms (inertial correction). The expansion in \eqref{eq:expn} is the weak-inertial correction that is used in a variety of problems to determine the influence of inertia in moderate $\Rey_l$ flows. 
Substituting \eqref{eq:expn} in the Navier-Stokes equation \eqref{eq:NS} gives the governing equation for the zeroth and first order problems,
\begin{subequations}
    \begin{align}
    \mathcal{O}(1):& \qquad \pderiv[2]{\ub_0}{\bar{z}}=\bar{\bnabla} \bar{p}_0, \label{eq:NS0} \\
    \mathcal{O}(\epsilon):& \qquad \pderiv[2]{\ub_1}{\bar{z}}=\bar{\bnabla} \bar{p}_1 +  (\ub_0 \cdot \bar{\bnabla}) \ub_0. \label{eq:NS1}
\end{align}
\end{subequations}
Next, we use the expansion \eqref{eq:expn} in the boundary conditions. It is straightforward to obtain the no-slip conditions at the respective order,
\begin{subequations}
    \begin{align}
        \mathcal{O}(1):& \qquad \ub_0 \vert_{\bar{z}=0}=\mathbf{0}, \label{eq:bc10}\\
        \mathcal{O}(\epsilon):& \qquad \ub_1\vert_{\bar{z}=0}=\mathbf{0} \label{eq:bc11}.
    \end{align}
\end{subequations}
To write the free-shear boundary condition at the liquid interface $\bar{z}=\bar{h}=\bar{h}_0+\epsilon \bar{h}_1$, we use the \emph{method of domain perturbations} \cite{Hinch91}. This method involves expanding a given function in Taylor series near the perturbed boundary. %$z=z_0+\epsilon z_1$, i.e. $f(z) = f(z)\vert_{z=z_0} + \epsilon z_1\; f'(z)\vert_{z=z_0} + \mathcal{O}(\epsilon^2)$. The prime $(\bullet')$ here denotes the derivative of the function. 
Thus, at $z=\bar{h}_0 +\epsilon \bar{h}_1$, we have
\begin{align}
    \pard{\ub}{\bar{z}}\biggr\vert_{\bar{z}=\bar{h}} &= \pard{\ub}{\bar{z}}\biggr\vert_{\bar{z}=\bar{h}_0} + \epsilon \bar{h}_1\pardn{2}{\ub}{\bar{z}}\biggr\vert_{\bar{z}=\bar{h}_0}+ \mathcal{O}(\epsilon^2) \nonumber \\ &= \;\pard{\ub_0}{\bar{z}}\biggr\vert_{\bar{z}=\bar{h}_0} + \epsilon \biggr(\pard{\ub_1}{\bar{z}}+ \bar{h}_1\pardn{2}{\ub_0}{\bar{z}}\biggr)\biggr\vert_{\bar{z}=\bar{h}_0} + \mathcal{O}(\epsilon^2).
\end{align}
The final expression was obtained using the expansion \eqref{eq:expn}. The free-shear condition at the respective order is then,
\begin{subequations}
    \begin{align}
    \mathcal{O}(1):& \qquad \pard{\ub_0}{\bar{z}}\biggr\vert_{\bar{z}=\bar{h}_0} = \mathbf{0}, \label{eq:bc20} \\
    \mathcal{O}(\epsilon):& \qquad \pard{\ub_1}{\bar{z}}\biggr\vert_{\bar{z}=\bar{h}_0} = -\bar{h}_1\; \pardn{2}{\ub_0}{\bar{z}}\biggr\vert_{\bar{z}=\bar{h}_0} = -\bar{h}_1 \bar{\bnabla} \bar{p}_0\vert_{\bar{z}=\bar{h}_0}. \label{eq:bc21}
\end{align}
\end{subequations}
We are finally at a stage where we can solve the governing equations and boundary conditions at their respective orders. At the zeroth order, we have the classical lubrication theory, which includes the Stokes equation \eqref{eq:NS0} and the boundary conditions \eqref{eq:bc10} and \eqref{eq:bc20}. Solving for $\ub_0$, we get  
\begin{align}
\ub_0 = \frac{\bar{z}}{2} (\bar{z}-2 \bar{h}_0) \bar{\bnabla} \bar{p}_0.
%\implies \ub_0 = \frac{1}{2}\Ca^{-1} \; z (2 h_0-z) \bnabla (\Delta h_0).
\label{eq:vel0_soln}
\end{align}
Next, we substitute \eqref{eq:vel0_soln} in the first-order governing equation \eqref{eq:NS1} and the boundary conditions \eqref{eq:bc11} and \eqref{eq:bc21}, and subsequently solve for $\ub_1$:
%\begin{align}
%    \pardn{2}{\ub_1}{z} = \bnabla p_1 + \frac{1}{4}z^2 (2h_0-z)^2 (\bnabla p_0 \cdot \bnabla) \bnabla p_0.
%\end{align}
%Solving the above equation for $\ub_1$ using the first order boundary conditions \eqref{eq:bc11} and \eqref{eq:bc21} gives, 
 \begin{align}
    \ub_1 = & \frac{\bar{z}}{2} (\bar{z}-2\bar{h}_0) \bar{\bnabla} \bar{p}_1 - \bar{z} \bar{h}_1 \bar{\bnabla} \bar{p}_0 + \frac{\bar{z}}{120}(\bar{z}^5-6\bar{h}_0\bar{z}^4+10 \bar{h}_0^2 \bar{z}^3-16 \bar{h}_0^5) \bar{\bnabla} \bar{p}_0 \cdot \bar{\bnabla} \bar{\bnabla} \bar{p}_0.
    \label{eq:vel1_soln}
\end{align}
Equation \eqref{eq:vel1_soln} introduces the inertial correction to the well-known Poiseuille flow profile of the Stokes solution \eqref{eq:vel0_soln}. One can factor out the $z$-dependence in these equations by computing the depth-averaged flow field, $\bar{\Ub}(x,y) = \bar{h}^{-1}\int_0^{\bar{h}} \ub \; \mathrm{d}\bar{z}$. Substituting \eqref{eq:expn} gives the depth-averaged flow-field of the form $\bar{\Ub}=\bar{\Ub}_0+ \epsilon \bar{\Ub}_1 + \mathcal{O}(\epsilon^2)$, where
\begin{align}
\bar{\Ub}_0 &= \frac{1}{\bar{h}_0} \int_0^{\bar{h}_0} \ub_0 \; \mathrm{d}\bar{z}, \\ \bar{\Ub}_1 &= \frac{1}{\bar{h}_0} \biggl(\int_0^{\bar{h}_0} \ub_1 \; \mathrm{d}\bar{z} + \frac{1}{\epsilon} \int_{\bar{h}_0}^{\bar{h}_0+\epsilon \bar{h}_1} \ub_0 \; \mathrm{d}\bar{z} - \frac{\bar{h}_1}{\bar{h}_0}\int_{0}^{\bar{h}_0} \ub_0 \; \mathrm{d}\bar{z} \biggr).
\end{align}
Applying \eqref{eq:vel0_soln} and \eqref{eq:vel1_soln} in the above integrals and evaluating, we obtain, 
\begin{align}
    \bar{\Ub}_0 = \frac{{-\bar{h}_0}^2}{3} \bar{\bnabla} \bar{p}_0, \quad
    \bar{\Ub}_1 = \frac{-\bar{h}_0^2}{3} \biggl(\bar{\bnabla} \bar{p}_1  + \frac{2 \bar{h}_1}{\bar{h}_0} \bar{\bnabla} \bar{p}_0 + \frac{54}{35}(\bar{\Ub}_0 \cdot \bar{\bnabla}) \bar{\Ub}_0 \biggr).
    \label{eq:U_temp}
\end{align}
We can simplify the above expressions by determining the pressure field. Noting that the pressure is independent of $\bar{z}$ (from \eqref{eq:PZ}), and is simply the dimensionless Laplace pressure,
\begin{equation}
    \bar{p} (\bar{x},\bar{y})= - \Ca^{-1} \delta^3 \bar{\Delta} \bar{h}(\bar{x},\bar{y}),
    \label{eq:press}
\end{equation}
where $\Ca= \eta V/ \gamma$ is the Capillary number. Since the pressure gradient has to balance the viscous stress at the leading order (see \eqref{eq:NS0}), we have $\delta \sim \Ca^{1/3}$. Disregarding the proportionality constant as done in the classical Stokes flow problem \cite{Limat04,Snoeijer05}, we get
\begin{equation}
    \delta=\Ca^{1/3} \implies \epsilon = \Ca^{2/3}\;\Rey_l.
    \label{eq:eps}
\end{equation}
Thus, by substituting \eqref{eq:expn} in \eqref{eq:press} and collecting terms of the same asymptotic order, we get
\begin{subequations}
    \begin{align}
    \mathcal{O}(1):&\quad \bar{p}_0 (\bar{x},\bar{y})= - \bar{\Delta} \bar{h}_0, \label{eq:press0}\\ \mathcal{O}(\epsilon):&\quad \bar{p}_1 (\bar{x},\bar{y})= - \bar{\Delta} \bar{h}_1.
    \label{eq:press1}
\end{align} 
\end{subequations}
The above expressions can be used in \eqref{eq:U_temp} to write $\bar{\Ub}_0$ and $\bar{\Ub}_1$ explicitly in terms of $\bar{h}_0$ and $\bar{h}_1$ alone,
\begin{subequations}
    \begin{align}
        \bar{\Ub}_0 &= \frac{{\bar{h}_0}^2}{3} \bar{\bnabla} ( \bar{\Delta} \bar{h}_0), \label{U0} \\
        \bar{\Ub}_1 &= \frac{\bar{h}_0^2}{3} \biggl(\bar{\bnabla} ( \bar{\Delta} \bar{h}_1)  + \frac{2 \bar{h}_1}{\bar{h}_0} \bar{\bnabla} (\bar{\Delta} \bar{h}_0) - \frac{54}{35}(\bar{\Ub}_0 \cdot \bar{\bnabla}) \bar{\Ub}_0 \biggr).
        \label{U1}
    \end{align}
\end{subequations}

\subsection{Mass conservation}
\label{sec2b}
In addition to the momentum conservation described in \S\ref{sec2a}, it is necessary for the flow-field to satisfy mass conservation given by,
\begin{equation}
\pderiv{\bar{h}}{\bar{t}} + \bar{\bnabla} \cdot (\bar{h} \bar{\Ub}) = 0.
\end{equation}
In the laboratory frame-of-reference, a drop moving in the positive $x$-axis with a (dimensionless) unit velocity has $\partial \bar{h}/\partial \bar{t} = - \partial \bar{h}/\partial \bar{x}$. Using this relation and the expansion \eqref{eq:expn}, we obtain the kinematic equation of the moving interface at each order,
\begin{subequations}
    \begin{align}
        \mathcal{O}(1):& \qquad -\pderiv{\bar{h}_0}{\bar{x}} + \bar{\bnabla} \cdot (\bar{h}_0 \bar{\Ub}_0) = 0  \label{cont00}\\
        \mathcal{O}(\epsilon):& \qquad -\pderiv{\bar{h}_1}{\bar{x}} + \bar{\bnabla} \cdot (\bar{h}_0 \bar{\Ub}_1 + \bar{h}_1 \bar{\Ub}_0) = 0
        \label{cont1}
    \end{align}
\end{subequations}
%We now seek to provide an inertial correction to the Stokes flow field. An Oseen-Poisuelle type weak-inertial correction is introduced here to the lubrication equation. As expected, Eq \eqref{cont00} is simply the continuity equation for Stokes flow (see Eqs \eqref{cont0}, \eqref{h0}), and the method to solve for $h_0$ was briefed in the previous Section. Eq \eqref{cont1} is the continuity equation for inertial correction flow, which we shall use to solve for $h_1$. Thus, we have, in dimensionless form,

%The advective term can be expanded with the help of Eq \eqref{U0_final} as
%\begin{equation}
%(\Ub_0 \cdot \bnabla) \Ub_0 = \frac{H^3}{x} (-\zeta f_x + f_y) (2 H' \fb + H \fb')
%\end{equation}
Note that $\bar{\Ub}_0$, $\bar{\Ub}_1$, $\bar{h}_0$ and $\bar{h}_1$ are yet unknown. However, since these fields are related through \eqref{U0} and \eqref{U1}, we need only solve for the latter two.

\section{Determining the interface profile}\label{sec3}

\subsection{Zeroth order solution: Stokes flow}
\label{subsec:Stokes}

The shape of the interface in the Stokes regime was determined in \citet{Limat04}. We briefly re-derive their results here for completeness. The kinematic equation for interface at $\mathcal{O}(1)$ is given by \eqref{cont00}. Using \eqref{U0} in \eqref{cont00}, 
\begin{align}
-3\;\pderiv{\bar{h}_0}{\bar{x}} + \bnabla \cdot (\bar{h}_0^3 \bar{\bnabla} (\bar{\Delta} \bar{h}_0))=0.
\label{h0}
\end{align}
%By determining $h_0(x,y)$ from \eqref{h0}, one can compute the velocity field near the corner by using the expression in \eqref{U0}. Thus, \eqref{U0} and \eqref{h0} provide the flow field and the interface profile respectively at the cornered contact line. 
Using a self-similarity ansatz, the solution is of the form
\begin{equation}
\bar{h}_0(\bar{x},\zeta) = \bar{x}\;H(\zeta),
\label{ss0}
\end{equation}
where $\zeta = y/x\;$ \cite{Limat04}. Substituting \eqref{ss0} in \eqref{h0}, one arrives at the ordinary differential equation,
\begin{align}
\zeta H' - H + H^3\;(f_x-\zeta {f_x}'+ {f_y}') + 3 H^2 H' (-\zeta f_x + f_y) =0
\label{H_eq}
\end{align}
where the prime ($\bullet'$) denotes derivative with respect to $\zeta$. The components of the vector $\fb(\zeta) = (f_x(\zeta), f_y(\zeta))$ are given by
\begin{subequations}
    \begin{align}
    f_x(\zeta) &= -\frac{1}{3}\biggl((1+3\zeta^2) H'' + \zeta (1+\zeta^2) H''' \biggr), \label{fx} \\ f_y(\zeta) &= \frac{1}{3} \biggl(2\zeta H'' + (1+\zeta^2) H'''\biggr).
    \label{fy}
    \end{align}
\end{subequations}

Eq.\eqref{H_eq} is a boundary value problem (BVP) that requires four boundary conditions to obtain a unique solution for $H(\zeta)$. By recognizing that the drop is symmetrical about its center line ($\zeta=0$), we get the first two conditions viz. $H'(0)=H'''(0)=0$. Next, we have at the contact lines $\zeta=\zeta_c$ (given), the height of the drop is zero i.e. $H(\zeta_c)=0$. The final condition requires no flux penetrating the contact line i.e., the net flux through any given cross section of the drop should be zero. $H(0)$ can be determined iteratively such that the no-flux condition is satisfied. However, the solution is extremely stiff near $\zeta=\zeta_c$ and convergence of the BVP would entail using special routines.

\begin{figure}
\centering
\begin{tabular}{cc}
\includegraphics[width=0.42\columnwidth]{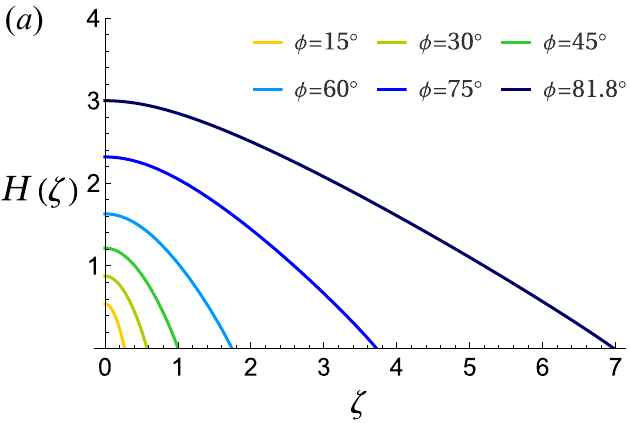} \hspace{1.5em} & \includegraphics[width=0.42\columnwidth]{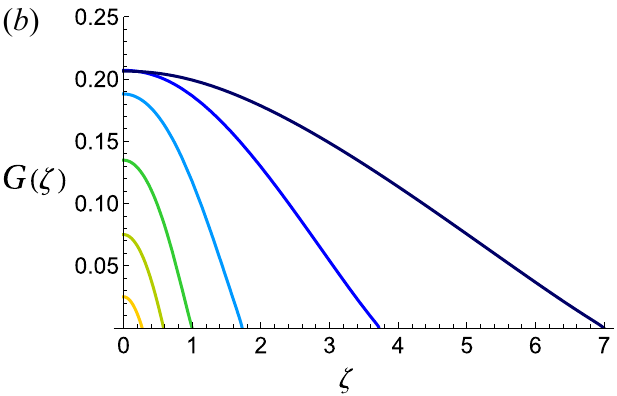}
\end{tabular}
\caption{Valid solutions of ($a$) Eq \eqref{H_eq} and ($b$) Eq \eqref{eq:G} for various corner semi-opening angle, $\phi$. For small $\phi$, the curves can be approximated as parabolas (see Appendix \ref{app:small}). Note that the plots are symmetrical about the $y$-axis for negative values of $\zeta$.}
\label{fig:H}
\end{figure}

So instead, we use an iterative procedure, following \citet{Limat04} and solve for an initial value problem (IVP). The value of $H(0)$ is first arbitrarily assigned. Note however that since $H(\zeta)$ relates directly to the height of the liquid interface, negative values are physically irrelevant and hence, excluded from the choice of $H(0)$. Next, we have $H'(0)=H'''(0)=0$ as before due to symmetry. Finally, based on a particular choice of $H''(0)$, one can solve the IVP to obtain multiple solutions for $H(\zeta)$, out of which some do not have a root (see Fig \ref{fig:flux_H}a); In other words, the drop interface does not meet the solid and form a contact line. These are not physically relevant solutions and may be discarded. However, there also exist multiple solutions of $H(\zeta)$ which have roots. The correct choice of $H''(0)$ is then the one that also obeys the no-flux condition at the contact line. Some of these valid solutions of $H(\zeta)$ for the initial conditions $H(0)$ and $H''(0)$ tabulated in Table \ref{TableH} are shown in Fig \ref{fig:H}(a); they yield unique values of $\zeta_c$. For example, when $H(0)=1.215$, the choice $H''(0)=-2.623$ satisfies the zero net flux condition as shown in Fig \ref{fig:flux_H}(b,c); This solution places the contact line along $\zeta=\zeta_c=1$ (up to 4 digits of precision) which corresponds to an opening angle $\phi = \tan^{-1}(\zeta_c)=45^\circ$. %Alternatively, for a predefined contact line $\zeta=\zeta_c$, one could have solved an equivalent boundary value problem (BVP), with $H(\zeta_c)=0$ as an imposed boundary condition; $H'(0)=H'''(0)=0$ as before due to reasons of symmetry. The appropriate choice of $H(0)$ should then be made which satisfies the zero net flux condition. However, a shooting method which is generally employed to solve the BVP, is computationally more cumbersome than the original IVP. Additionally, the solution is extremely stiff near $\zeta=\zeta_c$ and convergence of the BVP would entail using special routines.

\subsection{First order solution: Inertial correction}
\label{sec3b}

To determine the leading-order inertial correction, we use the mass conservation equation at $\mathcal{O}(\epsilon)$. Substituting \eqref{U0} and \eqref{U1} in \eqref{cont1}, we obtain the differential equation for the correction to the interface profile, $\bar{h}_1$,
\begin{align}
-3 \; \pderiv{\bar{h}_1}{\bar{x}} + \bar{\bnabla} \cdot \left(\bar{h}_0^3 \bar{\bnabla}(\bar{\Delta} \bar{h}_1) + 3 \bar{h}_0^2 \bar{h}_1 \bar{\bnabla}(\bar{\Delta} \bar{h}_0) - \frac{54}{35} \; \bar{h}_0^3 (\bar{\Ub}_0 \cdot \bar{\bnabla}) \bar{\Ub}_0 \right) = 0.
\label{h1}
\end{align}
with $\Ub_0$ given in \eqref{U0}. Keeping in mind \eqref{ss0}, a quick examination of this equation reveals that it admits a self-similar solution of the form
\begin{equation}
\bar{h}_1(\bar{x},\zeta) = \bar{x}^2 \;G(\zeta)  
\label{ss1}
\end{equation}
with $\zeta=y/x$ and $G(\zeta)$ a yet unknown function. Substituting the expressions for $h_0$ and $h_1$ from \eqref{ss0} and \eqref{ss1} respectively in \eqref{h1}, we arrive at an equation for $G(\zeta)$,
    \begin{align}
    \zeta G' & - 2 G + 2 H^2 (g_x + G f_x) - \zeta (H^2 (g_x + G f_x))' + (H^2 (g_y + G f_y))' = 0,
    \label{eq:G}
\end{align}
where the vector $\fb(\zeta) = (f_x(\zeta), f_y(\zeta))$ is given in \eqref{fx}-\eqref{fy} and $\gb(\zeta) = (g_x(\zeta), g_y(\zeta))$  is
\begin{subequations}
\begin{align}
g_x(\zeta) = \frac{1}{3} \biggl(6 G f_x + \frac{54}{35} H^4 (\zeta f_x - f_y) (2 H' f_x + H {f_x}') - \zeta (1+ \zeta^2) H G'''\biggr), \label{gxgya}
\end{align}
\vspace{-2em}
\begin{align}
g_y(\zeta) = \frac{1}{3} \biggl(6 G f_y  + \frac{54}{35} H^4 (\zeta f_x - f_y) (2 H' f_y + H {f_y}') + (1+ \zeta^2) H G'''\biggr). \label{gxgyb}
\end{align}
\end{subequations}

Note that $H(\zeta)$ is the Stokes solution for a contact line at some $\zeta=\zeta_c$, and is known at this stage from \S \ref{subsec:Stokes}. Eq.\eqref{eq:G} has to be supplemented with four boundary conditions to obtain a unique solution for $G(\zeta)$.  The symmetry of the drop automatically lets us assign $G'(0)=0$ and $G'''(0)=0$. Next, we note that the height of the interface should be zero at the contact line $\zeta=\zeta_c$ and this constraint sets $G(\zeta_c) =0$. The final boundary condition is obtained by specifying the value of $G''(0)$ (or $G(0)$); However, choosing any arbitrary value is not sufficient. Similar to the Stokes flow problem, the correct choice should result in a net zero-flux at any point on the contact line and through any cross-section of the drop (see Appendix \ref{app:flux}). The unique solutions of $G(\zeta)$ for a few specified values of $\zeta_c$ (or equivalently, the opening angle $\phi$) are shown in Fig.\ref{fig:H}(b); the corresponding choice of values of $G''(0)$ are also listed in appendix Table \ref{TableH}. In the example considered in \S \ref{subsec:Stokes} (drop having opening angle $\phi=45^\circ \; (\zeta_c=1)$), the correct choice of $G''(0)=-0.29$.

There are a few observations that need to be made here: First, $G(\zeta)$ directly relates to the \emph{correction} of the interface height due to inertia and so, it is allowed to assume negative values. Even so, it is interesting to note that the valid solution is always positive for the wide range of opening angles shown here. One can analytically show that it is always positive in the asymptotic limit of small opening angles (see Appendix \ref{app:small}). Second, $G(0)$ increases with the opening angle until about $\phi\approx 75^\circ$, beyond which it seems to saturate, rendering the solution sensitive to the choice of $G(0)$. For this reason, opting $G''(0)$ as the boundary condition is a more reliable alternative. %Finally, note that we solve a BVP here, prescribed by the conditions at $\zeta=0$ and $\zeta=\zeta_c$. But solving a BVP is computationally more expensive than an equivalent IVP, especially because the solution is extremely stiff near $\zeta_c$; 
We provide the approximate values of $G(0)$ and $G''(0)$ for a few opening angles in Table \ref{TableH}.

\section{Results}\label{sec4}

We can now use the solutions for $H(\zeta)$ and $G(\zeta)$ obtained in \S \ref{sec3} for various corner opening angles to determine the weakly-inertial interface profile and flow-fields. Inspired by the experiments of \citet{Puthenveettil13}, we shall use the properties of water and mercury as examples to quantify our findings because the drops of these liquids move in the inertial regime when their receding contact line destabilizes. At room temperature, the (density, surface tension, dynamic viscosity) of water and mercury are ($\rho_w=10^3\; \mathrm{kg\;m^{-3}}$; $\gamma_w=7 \times 10^{-2}\; \mathrm{N\; m^{-1}}$; $\eta_w=10^{-3}\;\rm{Pa\; s}$) and ($\rho_m=10^4\; \mathrm{kg\;m^{-3}}$; $\gamma_m=5\times 10^{-1}\;\mathrm{N\;m^{-1}}$; $\eta_m=1.5\times 10^{-3}\;\rm{Pa\;s}$) respectively. Water and mercury have critical Capillary numbers of $\Ca_{cr} \approx 4 \times 10^{-3}$ and $\Ca_{cr} \approx 1.5 \times 10^{-3}$\;\; \cite{Puthenveettil13}. 

Before we proceed, it is useful to define a local Reynolds number based on the dimensional distance $x$ from the corner, given by $\Rey_{x} = xV\rho/\eta = \bar{x}\; \Rey_l$. Also note that one can relate $\Rey_x$ to the capillary number via $\Rey_x = \Ca\; x/l_{vc}$, where $l_{vc}=\eta^2/(\gamma \rho)$ is the visco-capillary length of the liquid; For mercury and water, $l_{vc} \sim 10^{-4}\mu\mathrm{m}$ and $l_{vc} \sim 10^{-2}\mu\mathrm{m}$ respectively.

\subsection{Interface profile}

The liquid interface is obtained by substituting the expressions for $h_0$ and $h_1$ from \eqref{ss0} and \eqref{ss1}, 
\begin{align}
    \bar{h}(\bar{x},\zeta) & = \bar{x}\; H(\zeta) + \epsilon \bar{x}^2\; G(\zeta) + \mathcal{O}(\epsilon^2).
\end{align}
Truncating at $\mathcal{O}(\epsilon)$ and writing in dimensional form by using \eqref{eq:eps}, we get
\begin{align}
h(x,\zeta) & = \Ca^{1/3}\; x\; H(\zeta) + \Ca\;\Rey_x\; x \; G(\zeta).
\label{eq:h_final}
\end{align}
Note that $\Ca\; \Rey_x$ together is the local Weber number. The first term in \eqref{eq:h_final} is the Stokes solution, with the classical $\Ca^{1/3}$ scaling for partially-wetting liquids \cite{Cox86,Limat04}. By rewriting $\Rey_x = \Ca\;x/l_{vc}$, we see that the leading-order inertial term obeys a $\Ca^2$ scaling. %Comparing the terms, we see that Eq.\eqref{eq:h_final} is valid in the asymptotic limit $\Ca^{2/3} \Rey_x \ll 1$.

\begin{figure}[ht]
 \centering
 %\begin{tabular}{l}
 \includegraphics[width=0.45\columnwidth]{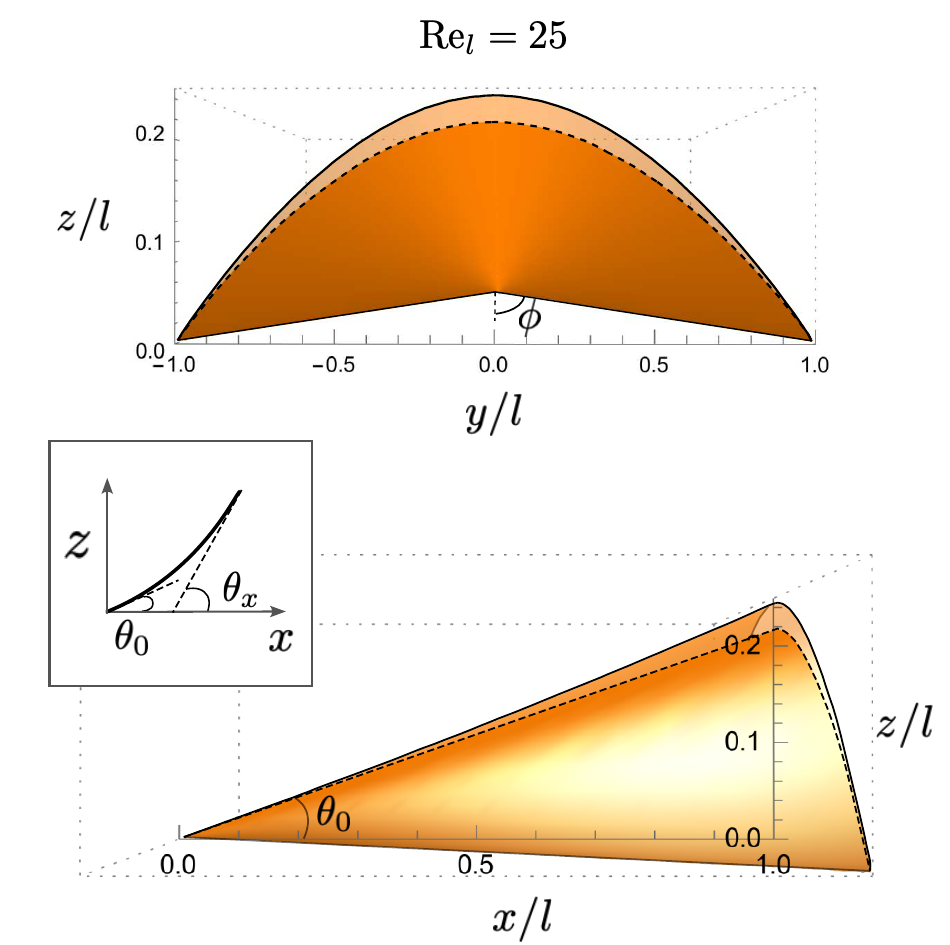} \\ %\hspace{-3em}
 %\includegraphics[width=0.9\columnwidth]{figs/dropside3.pdf}
 %\end{tabular}
 \caption{Front and side profiles of the liquid interface for a semi-opening angle of $\phi=45^\circ\;(\zeta_c=1)$, $\Ca=5.65\times 10^{-3}$ and $\Rey_l=25$. The liquid is assumed to have properties of water ($\Ca_\mathrm{cr}=4\times 10^{-3}$). The Stokes solution (cone) is shown by the dotted line while the inertia-corrected solution (cusp) is shown by the solid line. Inset: a schematic of the side angle $\theta_x$; the true side angle, $\theta_0$, computed at $x=0$ is the Stokes solution.}
 \label{fig:profile}
\end{figure}

We have seen in \S \ref{sec3b} that the solutions which satisfy the no-flux condition have $G(\zeta) \geq 0$ for all opening angles. Thus, the inertial correction in \eqref{eq:h_final} is always positive, which implies that the height of every point of the Stokes interface is increased due to inertia. A similar enhancement in the height of the liquid interface because of inertia has been reported previously for the Landau-Levich (fully-wetting) problem \cite{deRyck98}. In the Stokes limit, the interface height scales linearly with $x$, giving it a conical structure, as shown by the dashed lines in Figure \ref{fig:profile}. Since $\Rey_x$ varies linearly with $x$, the inertial correction in \eqref{eq:h_final} has a quadratic dependence with $x$. Indeed, we see its implication in the example shown in Figure \ref{fig:profile}, where the resulting surface has a cusp-like geometry.
%Furthermore, \eqref{eq:h_final} has a quadratic dependence on $x$ resulting from the inertial term, which physically implies the deformation of an otherwise conical interface into a cusp. 
Formation of cusps have also been reported in experiments at high contact line speeds, indicating that inertia might be playing an important role in the corner-to-cusp transition of moving contact lines \cite{Podgorski01,LeGrand05,Puthenveettil13}.

\begin{figure}[htbp]
\centering
\begin{tabular}{cc}
 \multicolumn{2}{c}{\includegraphics[width=0.55\columnwidth]{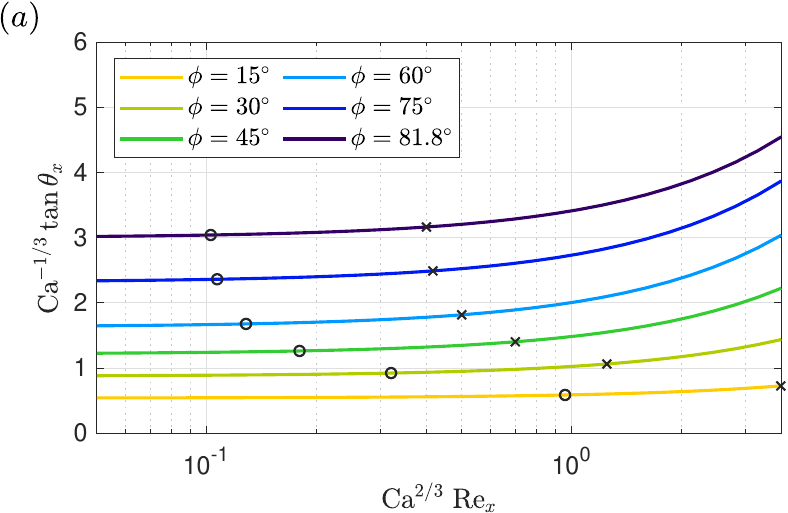}} \vspace{1em} \\
 \includegraphics[width=0.3\columnwidth]{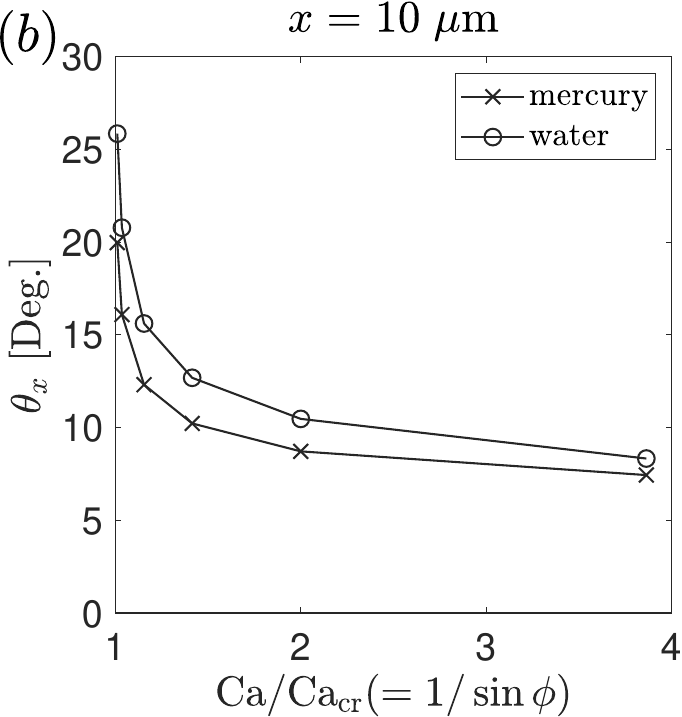} \hspace{1.5em} & \includegraphics[width=0.31\columnwidth]{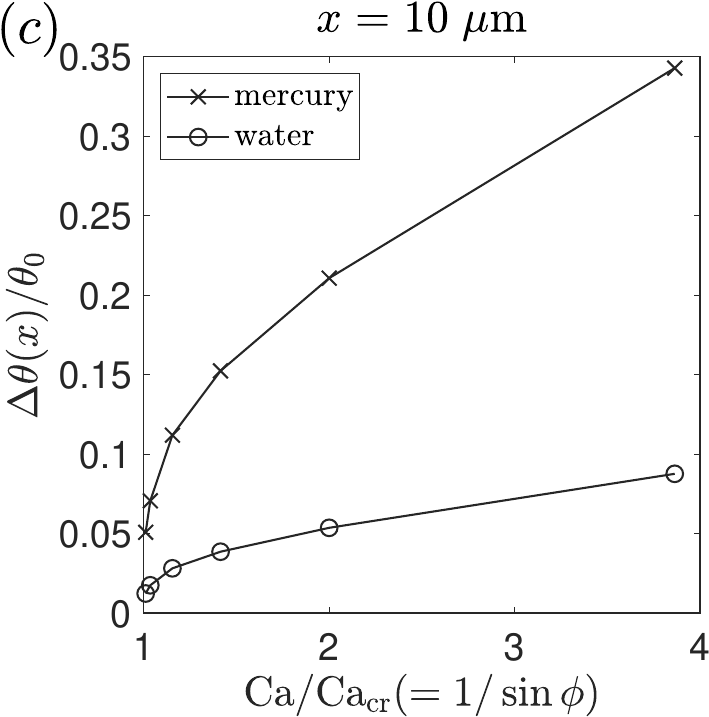}
 \end{tabular}
%\begin{tabular}{ll}
%\includegraphics[width=0.47\columnwidth]{figs/3dcl_th_abs.eps} & \includegraphics[width=0.48\columnwidth]{figs/3dcl_th_rel.eps}
%\end{tabular}
\caption{($a$) Variation of the side angle $\theta_x$ due to inertia given by \eqref{eq:theta_app} for various corner semi-opening angles. The cross and circle markers denote the values that get selected when using contact line dynamics \eqref{eq:phen} for mercury and water respectively at $x=10\; \mu\mathrm{m}$. ($b$) Same as markers in (a) but plotted solely against $\Ca$. ($c$) The fractional correction $\Delta \theta_x/\theta_0 = (\theta_x(x)-\theta_0)/\theta_0$ in the side angle due to inertia.}
\label{fig:side_angle}
\end{figure}

A useful quantification is the side angle of the drop, $\theta_x$. It is the internal angle in the $xz$ plane defined at a distance $x$ along the centerline ($\zeta=0$). A schematic of this side angle is shown in Figure \ref{fig:profile} (inset). It is formally defined as,
\begin{equation}
\tan \theta_{x} = \pderiv{h}{x}\biggr\vert_{\zeta=0} = \Ca^{1/3} H(0) + 2 \Ca\; \Rey_{x} G(0).
\label{eq:theta_app}
\end{equation} 
For a given semi-opening angle $\phi$, the side angle is a constant $\theta_0$ in the case of Stokes flow. However, by including inertial effects, we see that the side angle $\theta_x$ increases with the distance from the corner as well. To evaluate the side angle using \eqref{eq:theta_app}, one can utilize the Table \ref{TableH} which lists the values of $H(0)$ and $G(0)$ for different opening angles, $\phi$. Figure \ref{fig:side_angle}(a) shows this variation for various opening angles, $\phi$.

For small opening angles ($\phi \to 0$), one can show that the cross-section profile of the interface is parabolic, similar to the non-inertial case (see Appendix \ref{app:small}). Furthermore, in this case, we have the approximation $H(0)^3 \sim \tan^2 \phi $ and $G(0) \sim \tan^2 \phi$, giving us the $\theta-\phi$ relation in \eqref{eq:theta_phi_rel1}, valid for small $\phi$.  
%\begin{align}
%    \tan \theta_x \approx \left(\frac{35}{16} \right)^{1/3} \Ca^{1/3} \tan^{2/3}\phi + \frac{140}{143}\; \Ca\;\Rey_x \tan^2\phi 
%\end{align}
To facilitate comparison with other works in literature and experiments (see review \cite{Snoeijer11}), we take cube of this expression and truncate it at the leading-order in $\Ca^{2/3} \Rey_x (\ll 1)$ to get the inertia-corrected relation,
\begin{equation}
    \tan^3 \theta_x \approx \frac{35}{16} \Ca \tan^2 \phi \left( 1 + 2.262\; \Ca^{2/3} \Rey_x \tan^{4/3} \phi \right) .
\end{equation}

Up until this point, we have not included any contact line dynamics in our model; the opening angle $\phi$ (or equivalently, $\zeta_c$) was chosen independently. However, it has been observed in experiments that $\phi$ reduces with increasing $\Ca$ such that the velocity of the contact line is maintained at the critical value \cite{Blake79,Petrov85}. We account for this variation by including the commonly-used phenomenological relation, 
\begin{equation}
    \sin \phi = \Ca_\mathrm{cr}/\Ca,
    \label{eq:phen}
\end{equation}
in our model \cite{Blake79,Petrov85}; A more detailed mechanism for selection of $\phi$ was explored in \citet{Snoeijer07}. While it is natural to expect inertia to increase with the drop velocity, one should keep in mind that the competing viscous forces also increase due to the corresponding narrowing of the corner. To analyze the non-trivial effects that arise from introducing the contact line dynamics in the model, we shall use mercury and water at room temperature as example liquids. The properties of mercury and water were described at the beginning of this section. For any specified liquid, since $\phi$ is enslaved to $\Ca$, there is a unique value of $\theta_x$ for a given $\Ca$ and distance $x$ from the corner. These values for mercury and water at $x=10\;\mu\mathrm{m}$ are highlighted in Figure \ref{fig:side_angle}(a) by crosses and circles respectively. Since $\Rey_x = \Ca\;x/l_{vc}$, $\theta_x$ can be expressed entirely in terms of $\Ca$ after inclusion of contact line dynamics, as shown in Figure \ref{fig:side_angle}(b) for some fixed $x\;(=10\;\mu\mathrm{m})$. Contrary to what one might expect from \eqref{eq:theta_app}, we see that $\theta_x$ reduces with $\Ca$, consistent with experimental observations \cite{LeGrand05,Puthenveettil13}. This is due to the stronger influence of the narrowing geometry (decreasing $H(0)$ and $G(0)$ values) compared to increasing $\Ca$. Despite this, we see in Figure \ref{fig:side_angle}(c) that the fractional correction introduced by inertia increases with $\Ca$. Also note that the correction for mercury is higher than water in spite of the former's smaller $\Ca_\mathrm{cr}$, due to its much higher $\Rey_x$ values (at $x=10\mu\mathrm{m}$, $\Rey_x\sim \mathcal{O}(1)$ for water and $\Rey_x \sim \mathcal{O}(10)$ for mercury).

\subsection{Depth-averaged flow field}

Using the interface profile determined in the previous section, one can compute the pressure fields using \eqref{eq:press0}. At the zeroth and first orders, these are respectively (in dimensionless form),
\begin{align}
    \bar{p}_0 &= - \frac{(1+\zeta^2)}{\bar{x}} H''(\zeta), \label{eq:press_Stokes}\qquad \mathrm{and,} \\ \bar{p}_1 &= -(1+\zeta^2) G''(\zeta) + 2 \zeta G'(\zeta) - 2 G(\zeta). \label{eq:press_inertial}
\end{align}
Note that when approaching the corner, $\bar{x}\to 0$, the Stokes flow pressure $\bar{p}_0$ becomes singular, whereas the inertial correction $\bar{p}_1$ does not. This is unlike the case of a straight contact line, where the leading-order stress from inertia exhibits a logarithmic singularity \cite{Varma21}.

The depth-averaged velocity field at the zeroth order (Stokes flow) is obtained by substituting \eqref{ss0} in \eqref{U0},
\begin{equation}
    \bar{\Ub}_0(\zeta)=H(\zeta)^2 \fb(\zeta),
    \label{eq:U0_final}
\end{equation}
with $\fb(\zeta)$ given in \eqref{fx}-\eqref{fy}. Likewise, at the first order, we get by substituting \eqref{ss0}, \eqref{ss1} and \eqref{eq:U0_final} in \eqref{U1},
\begin{equation}
\bar{\Ub}_1(\bar{x},\zeta)= \bar{x}\; H(\zeta) \gb(\zeta),
\label{U1_final}
\end{equation}
with $\gb(\zeta)$ given in \eqref{gxgya}-\eqref{gxgyb}. Thus, the total flow velocity is then $\bar{\Ub}(\bar{x},\zeta) = \bar{\Ub}_0(\zeta) + \epsilon \bar{\Ub}_1(\bar{x},\zeta) + \mathcal{O}(\epsilon^2)$. In dimensional form, this is
\begin{align}
\frac{\Ub(x,\zeta)}{V} & = {H(\zeta)}^2  \fb(\zeta) + \Ca^{2/3} \Rey_{x}\;H(\zeta) \gb(\zeta).
\label{eq:U_final}
\end{align}
We remind that $\Ub = (U_x,U_y)$ is the local depth-averaged velocity field within the liquid, and is different from the global velocity of the drop/contact line, $\boldsymbol{V}=(V,0)$. The Stokes velocity field depends only on the semi-opening angle, $\phi$, while the inertial correction depends on $\phi$, $\Ca$ and distance from the corner, $x$. Furthermore, since $\Rey_x = \Ca\;x/l_{vc}$, the leading-order inertial term scales as $\Ca^{5/3}$.
  
\begin{figure}
  \centering
 \begin{tabular}{c}
 \includegraphics[width=0.55\columnwidth]{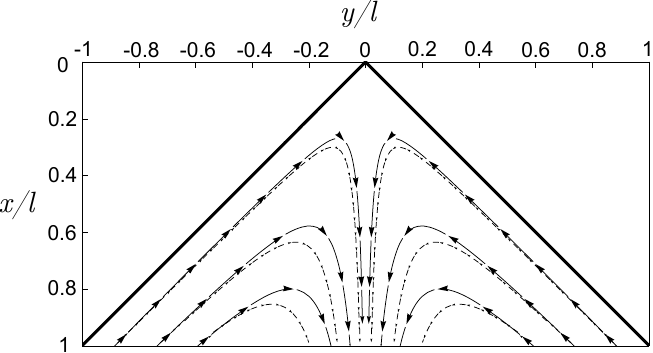}
 \end{tabular}
 \caption{Streamlines of the depth-averaged flow near the corner shown in the reference frame of the moving drop i.e. $(\Ub-\boldsymbol{V})/V$, obeying \eqref{eq:U_final}. The dotted lines are the Stokes solution while the solid lines are the inertia-corrected solution. A semi-opening angle of $\phi=45^\circ$ and $\Ca = 5.65 \times 10^{-3}$ is chosen to mimic water ($\Ca_\mathrm{cr}=4\times 10^{-3}$). The local Reynolds number at some $x=l$ is taken to be $\Rey_l=25$; For water, $l \approx 250\mu\mathrm{m}$.}
 \label{fig:streamlines}
\end{figure}

In Figure \ref{fig:streamlines}, we show as an example, the Stokes and inertia-corrected streamlines in the reference frame of a moving drop. One can make the following qualitative observation, true for all opening angles: The fluid approaching the corner is forced further towards it due to the inertia. After turning the corner, the fluid accelerates away from it, and at the same time, is drawn further towards the centerline of the drop  ($\zeta=0$). The flow velocity is maximum along this centerline. 

\begin{figure}[htbp]
  \centering
 \begin{tabular}{cc}
 \multicolumn{2}{c}{\includegraphics[width=0.55\columnwidth]{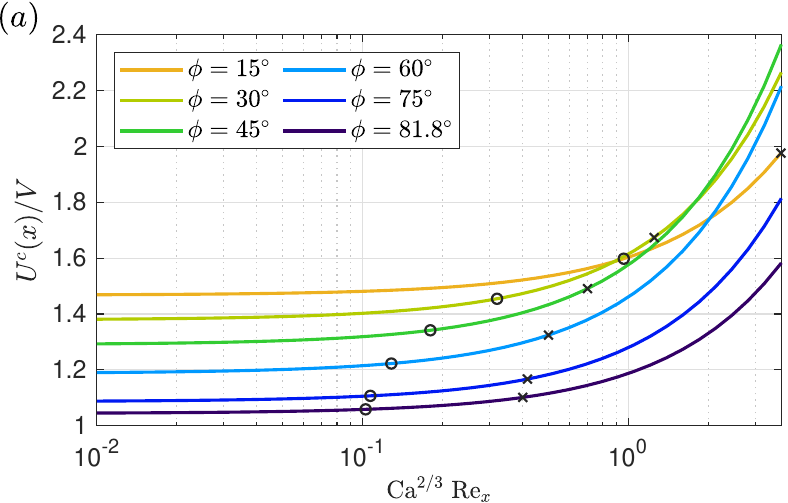}} \vspace{1em} \\
 \includegraphics[width=0.3\columnwidth]{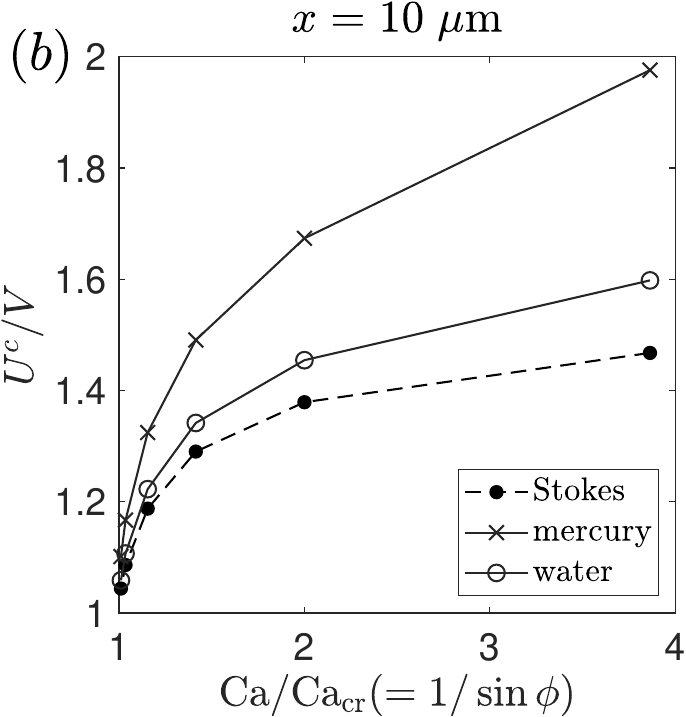} \hspace{1.5em} & \includegraphics[width=0.31\columnwidth]{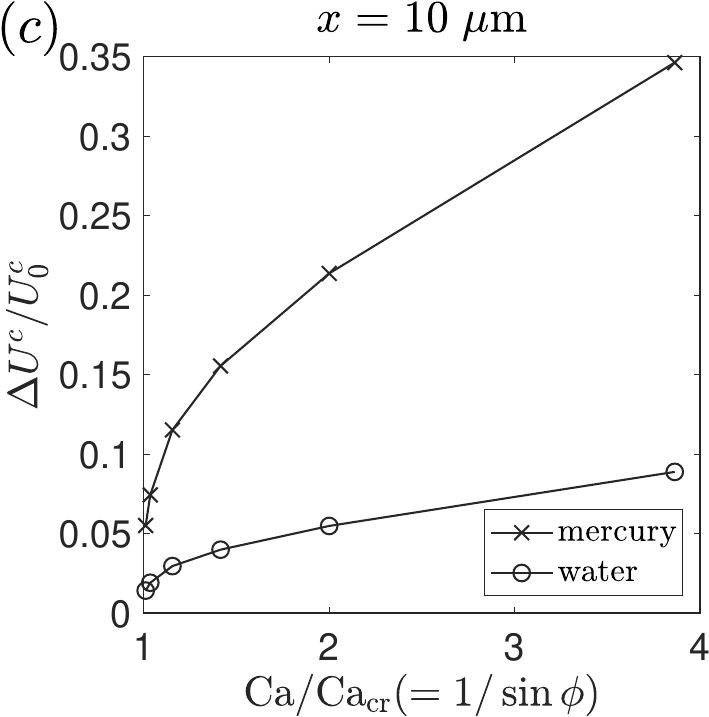}
 \end{tabular}
 \caption{($a$) Variation of the centerline velocity due to inertia given by \eqref{eq:U_final} for various corner semi-opening angles. The cross and circle markers denote the values that get selected when using contact line dynamics \eqref{eq:phen} for mercury and water respectively at $x=10\; \mu\mathrm{m}$. ($b$) Same as markers in (a) but plotted solely against $\Ca$. ($c$) The fractional correction $\Delta U^c/U^c_0 = (U^c(x)-U^c_0)/U^c_0$ in the centerline velocity due to inertia.}
 \label{fig:vel_scaling}
\end{figure}

To get a quantitative estimate of the effect of inertia, we compute the  centerline velocity, where the depth-averaged flow is strictly along the $x$-axis i.e. $U^c(x) = U_x(x,0)$. It is determined using \eqref{eq:U_final} where the values of $H(0), f_x(0)$ and $g_x(0)$ for a few opening angles can be read off from Table \ref{TableH}; Note that $f_y(0)$ and $g_y(0)$ are strictly zero. The results are shown in Figure \ref{fig:vel_scaling}(a) for various semi-opening angles, $\phi$.
One can see immediately that the centerline velocity is more than the drop velocity i.e. $U^c/V>1$, even in the Stokes limit ($\Rey_x\to 0$). With the inclusion of the inertial correction, this value increases with $\Rey_x$, for all opening angles. Thus, we come to the conclusion that for a given drop velocity, the flow velocity increases with the distance from the corner due to inertia. The magnitude of this increase (i.e the slope of the curve) depends on the opening angle, with the most increase seen for intermediate values.

We did not include any contact line dynamics for generating Figure \ref{fig:vel_scaling}(a), which is to say that $\phi$ and $\Ca$ were treated as independent parameters. To get a more practical picture, we shall now include the phenomenological relation between these two quantities given by \eqref{eq:phen}, as we did in the previous section for the side angle. The values of $U^c/V$ that get selected in this manner for two example liquids viz. mercury and water at $x=10\mu\mathrm{m}$ are shown with crosses and circles respectively. We show these explicitly in terms of the $\Ca$ in Figure \ref{fig:vel_scaling}(b). Notably, the Stokes centerline velocity, $U^c_0/V$, being a function only of $\phi$, is given by a universal curve for all liquids and at any distance from the corner. The inertia-corrected velocities for both mercury and water at some $x$($=10\mu\mathrm{m}$ here) are not only higher than the Stokes velocity, as mentioned, but it also increases with $\Ca$; In comparison, the centerline velocity of mercury is much higher than water at any distance from the corner because of the higher local Reynolds number of the former.
Figure \ref{fig:vel_scaling}(c) shows the fractional correction introduced by inertia at $x=10\mu\mathrm{m}$ which also increases with $\Ca$. This implies that inertia overcomes the enhanced viscous dissipation caused by the narrowing of the corner at higher speeds.

\section{Summary and discussions}\label{sec5}

We studied the effect of inertia in receding contact lines of partially-wetting liquids beyond their critical capillary number when they destabilize and form a corner. To this end, we modeled the flow-field near the contact line using thin-film equations, as is classically done, but retained the convective (inertial) term. The leading-order inertial contribution was then obtained by linearizing the equations using a regular perturbation expansion. The inertial effects were found to be significant when $\Ca^{2/3} \Rey_x \gtrsim 1$, where $\Rey_x$ is the local Reynolds number based on the distance from the corner, $x$. % or equivalently, when $x \gtrsim \Ca^{-5/3} l_{vc}$ where $l_{vc}=\eta^2/(\gamma \rho)$ is the visco-capillary length. 
For destabilized contact lines of many common liquids, including mercury and water considered here, $\Ca \sim 10^{-3}-10^{-2}$. %and $l_{vc} \sim 10^{-4} - 10^{-2}\;\mu\mathrm{m}$. 
This implies that inertia becomes significant when $\Rey_x \gtrsim 10-10^2$, which corresponds to experimentally relevant length scales of $x \sim 10-100\; \mu\mathrm{m}$ for mercury and $x \sim 100-1000\; \mu\mathrm{m}$ for water. Gravitational effects can be ignored as these distances fall well within their respective capillary lengths. %Thus, within $l_{vc} \ll x \ll l_c$, one can safely assume the interface to be slowly varying. Furthermore, we assumed the interface to be thin compared to its other dimensions, allowing us to use the lubrication approximation. 

In the Stokes regime, the thin-film equations predict the receding contact line to be of conical structure with a self-similar flow within \cite{Limat04,Snoeijer05}. Here, we have provided a self-similar leading-order inertial correction to this interface profile and the flow-field. We have also shown that for all corner opening angles, inertia enhances the interface height. A similar enhancement in the liquid film thickness by inertia was reported in the fully-wetting case (Landau-Levich problem)\cite{Koulago95,deRyck98}. Furthermore, the leading-order inertial correction was found to scale linearly with the local Weber number ($\We_x = \Ca\; \Rey_x$), giving rise to a cusp-like shape of the interface for all corner opening angles. However, note that in computing the correction, we assumed the contact line to remain a straight corner ($\phi$ independent of $x$), which is not true in general, as experiments report a corner-to-cusp transition of contact lines at high speeds \cite{LeGrand05,Puthenveettil13}.

To understand the inertial effects on the flow-field, we computed the depth-averaged velocity near the contact line. The inertial correction was found to scale as $\Ca^{5/3}$ at  given distance from the corner. We also found that inertia increases the fluid influx from the bulk and forces the flow further towards the corner. Consequently, this also enhances the outflux through the centerline of the drop for all corner opening angles. 
%For a given drop opening angle, the Stokes flow velocity is independent of $\Ca$, but the inertial correction was found to scale as $\Ca^{2/3}\; \Rey_x$. We quantified the inertial contribution by analyzing the free-shear velocity along the centerline of the drop, there is a higher flux through the centerline (see Appendix).Along with $\Ca$ and $\Rey_x$, the inertial contribution also depends on the corner geometry -- the  in our case. 

We then introduced contact line dynamics in our model using a simple phenomenological relation for the corner opening angle, viz. $\sin \phi = \Ca_\mathrm{cr}/ \Ca\;$ \cite{Blake79}, which is to  say that increasing $\Ca$ reduces the opening angle. In doing so,  we introduced a competition between the increased drop speed and the corner confinement on the flow; the former enhances the fluid inertia while the latter suppresses it. Interestingly, we found that despite this competition, there is an overall increase in the inertial contribution with increase in $\Ca$. We have shown this numerically by taking contact lines of water and mercury as examples. While these are theoretical estimates, note that in practice, the cornered end undergoes either a pearling instability or forms a rivulet when $\phi \ll 1$; In experiments, it happens around $\phi =\pi/6$ or $30^\circ\;$ \cite{Podgorski01,Puthenveettil13}. In future works, it would be interesting to see how inertia itself affects this instability \cite{Snoeijer07}. 

While the assumption of $\epsilon \;(=\Ca^{2/3} \Rey_l) \ll 1$ was used to obtain a perturbative correction to the Stokes solution, the asymptotic expansion is valid numerically even for $\epsilon \sim \mathcal{O}(1)$. For fast motion of contact lines where $\epsilon \gg 1$, the flow develops a viscous boundary layer close to the surface. In such cases, the use of boundary layer theory instead has been shown to agree well with experiments \cite{Puthenveettil13}. 

The Stokes flow stresses diverge as $1/x$ when approaching the corner, $x \to 0$ (see \eqref{eq:press_Stokes}). It has been observed in experiments that the corner rounds off around the visco-capillary length, $l_{vc}$, to regularize this corner singularity \cite{Peters09}. Additional contact line physics can be implemented in the model to capture this fine tip structure \cite{Peters09}, although no such regularization was used in the present work. This makes our analysis less suitable for comparison with experiments when $x \lesssim l_{vc}$. Furthermore, far away from the corner, our analysis is not valid close to and beyond the capillary length, $l_c$, where gravitational effects become important. Thus, it is best applicable in $l_{vc} \ll x \ll l_{c}$. 

The inertial correction provided in the present work cannot be applied to advancing contact lines because of the large capillary numbers at which destabilization occurs ($\Ca_\mathrm{cr} \sim \mathcal{O}(10)$), accompanied by a large interface deformation \cite{Snoeijer13}. However, it remains relevant for addressing problems concerning destabilized receding contact lines, both from a fundamental standpoint as well as in many industrial processes operating at moderately large Reynolds numbers \cite{Lohse22,Winkels11,Snoeijer13}. 

\begin{acknowledgments}
The author thanks Charu Datt for many valuable discussions. The author is also grateful to Baburaj Puthenveettil for discussions and encouragement, and to Christina Kurzthaler for providing a detailed feedback on the manuscript.
\end{acknowledgments}

%%%%%%%%%%%%%%%%
\appendix

\section{Values at centerline of the drop}

\begin{table}[h]
\small
\caption{The centerline values ($\zeta=0)$ for various semi-opening angles, $\phi$ (or equivalently, $\zeta_c(=\tan \phi)$) used in the main text.}
\label{TableH}
\begin{tabular*}{0.85\columnwidth}{@{\extracolsep{\fill}}lllllll}
$\phi\;[\mathrm{Deg.}],\;\zeta_c$ & $H(0)$ & $H''(0)$ & $G(0)$ &  $G''(0)$ & $f_x(0)$ & $g_x(0)$ \vspace{0.5em} \\
\hline
%15 & 0.54 & -15.1 & 0.013 & -0.3 & 5.033 & 0.0707 \\ without the 54/35
%30 & 0.876 & -5.39 & 0.046 & -0.28 & 1.797 & 0.1448\\
%45 & 1.215 & -2.62 & 0.087 & -0.19 & 0.874 & 0.1848\\
%60 & 1.63 & -1.34 & 0.12 & -0.09 & 0.447 & 0.1749\\
%75 & 2.319 & -0.606 & 0.134 & -0.03 & 0.202 & 0.1255\\
%81.8 & 3 & -0.348 & 0.1337 & -0.01 & 0.116 & 0.0931\\
$15,\; 0.27$ & 0.54 & -15.1 & 0.025 & -0.68 & 5.033 & 0.2517 \\
$30,\;0.58$ & 0.876 & -5.39 & 0.075 & -0.458 & 1.797 & 0.2695\\
$45,\;1$ & 1.215 & -2.62 & 0.1355 & -0.29 & 0.874 & 0.237\\
$60,\;1.73$ & 1.63 & -1.34 & 0.185 & -0.138 & 0.447 & 0.168\\
$75,\;3.73$ & 2.319 & -0.606 & 0.207 & -0.039 & 0.202 & 0.0836\\
$81.8,\;6.94$ & 3 & -0.348 & 0.2064 & -0.0136 & 0.116 & 0.0479\\
\hline
\end{tabular*}
\end{table}
 One can use this table, for example, to determine the magnitude of the centerline depth-averaged velocity of the drop given by $U^c/V = H(0)^2\; f_x(0) + \Ca^{2/3} \Rey_x\; H(0)g_x(0) $ or the side angle in \eqref{eq:theta_app}.

\section{Approximation for small opening angles}
\label{app:small}

\begin{figure}[t]
\centering
\begin{tabular}{cc}
 \includegraphics[width=0.31\columnwidth]{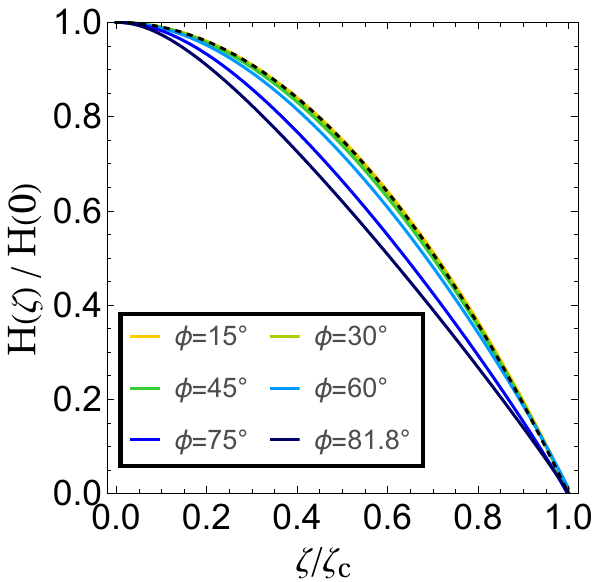} \hspace{1.5em} & \includegraphics[width=0.31\columnwidth]{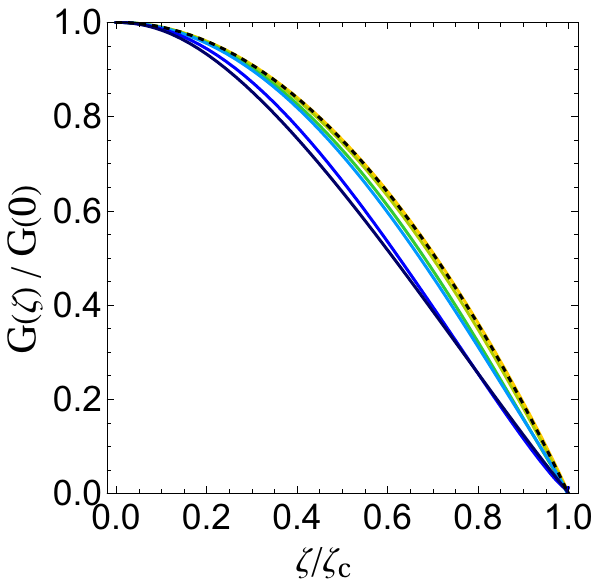}
 \end{tabular}
\caption{Rescaled solutions $H(\zeta)$ and $G(\zeta)$ from Figure \ref{fig:H} for various opening angles, $\phi$ (or equivalently $\zeta_c=\tan \phi$). Dashed line shows the parabolic curve $1-(\zeta/\zeta_c)^2$.}
\label{fig:supp}
\end{figure}

The functions $H(\zeta)$ and $G(\zeta)$ obtained in \S \ref{sec3} (Figure \ref{fig:H}) are rescaled with their values at the centerline ($\zeta=0$) in Figure \ref{fig:supp}. For small $\phi$ (equivalently, $\zeta_c \ll 1$), the solutions approximate as a single curve which is a function of $\zeta/\zeta_c$. Thus, in these cases, we can write $H(\zeta) = H(0) \hat{H}(\zeta/\zeta_c)$ and $G(\zeta) = G(0) \hat{G}(\zeta/\zeta_c)$, with the properties  $\hat{H}(\pm 1)=\hat{G}(\pm 1)=0,\; \hat{H}(0)=\hat{G}(0)=1$. In fact, from Figure \ref{fig:supp}, we see that the curve is approximately described by a parabola \emph{for both $\hat{H}$ and $\hat{G}$},
\begin{equation}
     \hat{H}(\zeta/\zeta_c) = \hat{G}(\zeta/\zeta_c) \approx 1-\left(\frac{\zeta}{\zeta_c}\right)^2.
     \label{eq:B1_approx}
\end{equation}
The former is the small-angle approximation given by \citet{Limat04}.

Using \eqref{eq:B1_approx} in \eqref{fx} and \eqref{fy}, we get
\begin{equation}
    f_x \approx \frac{2(1+3\zeta^2)}{3\zeta_c^2} H(0), \quad f_y \approx \frac{-4\zeta}{3\zeta_c^2} H(0)
    \label{eq:f_approx}
\end{equation}
respectively. By using \eqref{eq:B1_approx} and \eqref{eq:f_approx} in the expression for the Stokes flow flux \eqref{A} to satisfy the zero-net flux condition, we get the approximate expression for $H(0)$ given in \eqref{eq:H0}. Computing $g_x$ and $g_y$ (Eqns. \eqref{gxgya} and \eqref{gxgyb}) in a similar fashion, and using it in \eqref{A2}, we get the net inertial flux, which when equated to zero, gives the expression for $G(0)$ in \eqref{eq:G0}. Thus, for small opening angles i.e. $\zeta_c \ll 1$,
\begin{align}
    H(0) &\approx (35/16)^{2/3}\; \zeta_c^{2/3} \label{eq:H0} \\
    G(0) &\approx \frac{70}{143}\zeta_c^2 \frac{(1-9\zeta_c^2/17)}{(1+\zeta_c^2/2)}  \label{eq:G0} \\
    f_x(0) &\approx \frac{2}{3\zeta_c^2} H(0),\quad f_y(0) =0 \\
    g_x(0) &\approx \frac{4}{3\zeta_c^2} H(0) G(0),\quad g_y(0) =0
\end{align}
Since $G(0)$ in \eqref{eq:G0} is positive and $\hat{G}(\zeta/\zeta_c)$ is parabolic, the interface profile $G(\zeta)$ is always positive. 

Using the above approximation along with the fact that $\zeta_c=\tan\phi$, the $\theta-\phi$ relation in \eqref{eq:theta_app} becomes  
\begin{equation}
    \tan \theta_x \approx \left(\frac{35}{16} \right)^{1/3} \Ca^{1/3} \tan^{2/3}\phi + \frac{140}{143}\; \We_x \tan^2\phi
    \label{eq:theta_phi_rel1}
\end{equation}
where $\We_x = \Ca\; \Rey_x$ is the local Weber number. Note that the approximation of $G(0)$ in \eqref{eq:G0} is less accurate compared to $H(0)$ because of the existence of derivatives in the calculation of the flux.

\section{Flux near the contact line}
\label{app:flux}

\begin{figure}[htbp]
\centering
\begin{tabular}{ccc}
\includegraphics[width=0.3\textwidth]{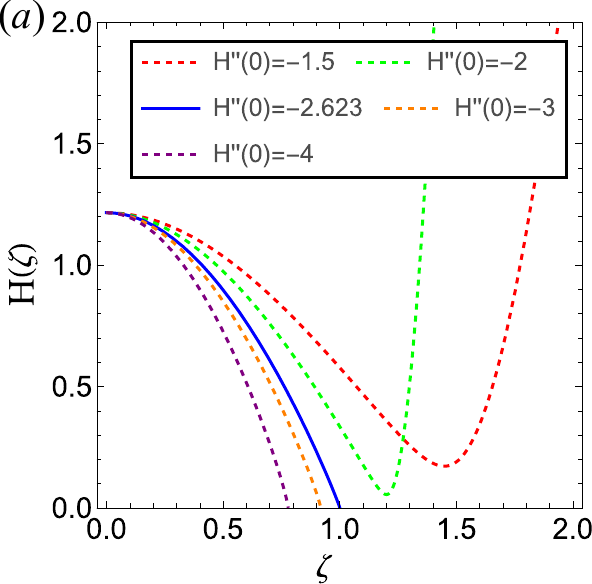} & \includegraphics[width=0.31\textwidth]{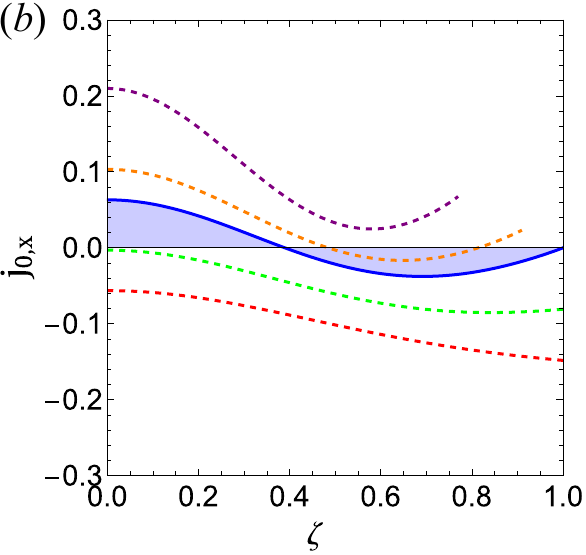} & \includegraphics[width=0.3\textwidth]{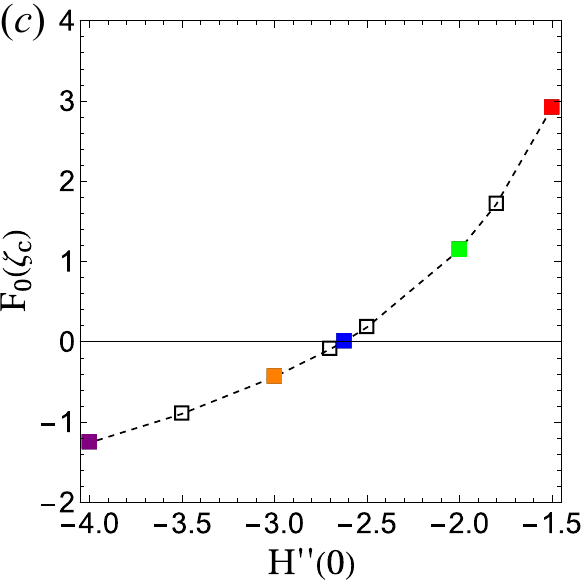} \\
\includegraphics[width=0.305\textwidth]{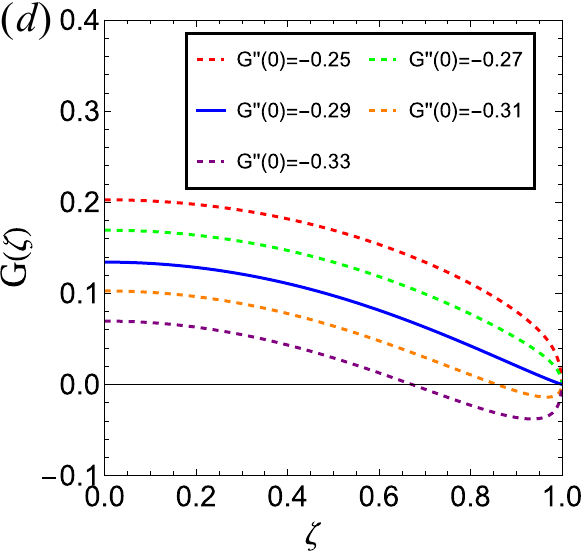} & \includegraphics[width=0.312\textwidth]{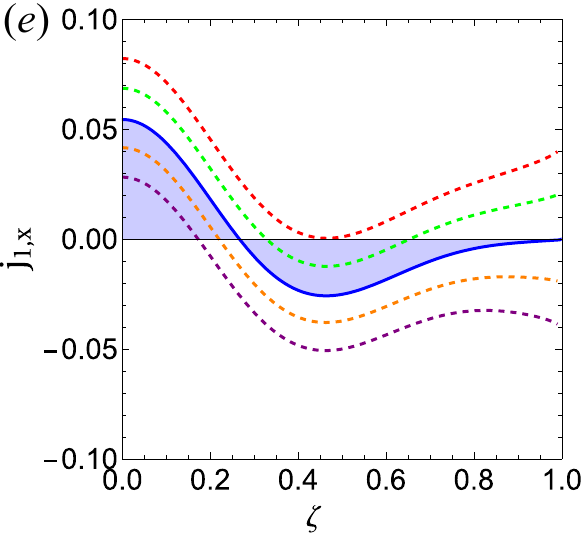} & \includegraphics[width=0.295\textwidth]{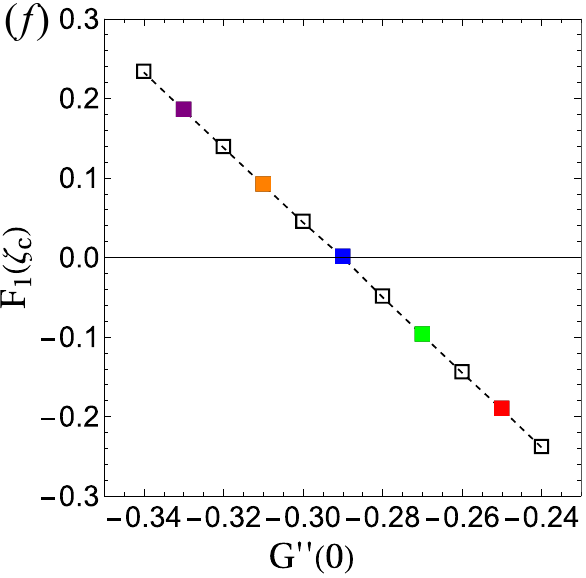} 
\end{tabular}
\caption{$\mathcal{O}(1)$ Stokes regime: ($a$) Solutions of \eqref{H_eq} with the boundary conditions $H(0)=1.215$, $H'(0)=H'''(0)=0$ and for different values of $H''(0)$. The physically valid solution (shown in blue solid line) is decided by the zero-net flux condition through the cross-section ($yz$ plane). ($b$) Local flux along a cross-section; Only $\zeta \geq 0$ is shown for reasons of symmetry. The area under each curve represents the net flux through the cross-section. ($c$) The net flux (Eq \ref{A}, shaded area in (b)) is proportional to $F_0(\zeta)$, which is shown here for the different values of $H''(0)$. A zero net flux i.e. $F_0(\zeta_c) = 0$, is obtained only for $H''(0)=-2.623$, making it the correct solution, and results in $\zeta_c=1$ or $\phi=45^\circ$. $\mathcal{O}(\epsilon)$ Inertial correction: ($d$) Solutions of $G(\zeta)$ in Eq \eqref{H_eq} with the boundary conditions $G(\zeta_c)=0$, $G'(0)=G'''(0)=0$ and for different values of $G''(0)$. The correct choice of $G''(0)$ is decided by the zero-net flux condition along any arbitrary cross-section of the drop. ($e$) Local inertial flux correction; For reasons of symmetry, only $\zeta \geq 0$ is shown. The area under each curve represents the net inertial flux through the cross-section. ($f$) The net inertial flux (Eq \ref{A2}) is proportional to $F_1(\zeta)$, which is shown here for the different values of $G''(0)$. A zero net flux i.e. $F_1(\zeta_c) = 0$, is obtained only for $G''(0)=-0.29$, making it the correct choice. }
\label{fig:flux_H}
\end{figure} 

$\bar{\Ub}$ is the depth-averaged velocity field rescaled by the drop velocity. The relative velocity is then $\bar{\Ub}-\hat{\xb}$, where $\hat{\xb}$ is the unit vector in the positive $x$ direction. The dimensionless local flux $j_x$ is then
\begin{equation}
\bar{j}_x = \bar{h}(\bar{\Ub} \cdot \hat{\xb}-1)
\end{equation}
Like in the rest of the article, expanding $\bar{h}=\bar{h}_0+ \epsilon \bar{h}_1$ and $\bar{\Ub} = \bar{\Ub}_0+ \epsilon \bar{\Ub}_1$ gives $\bar{j}_x = \bar{j}_{0,x} + \epsilon \bar{j}_{1,x} + \mathcal{O}(\epsilon^2)$ with,
\begin{align}
\bar{j}_{0,x} = \bar{h}_0 (\bar{U}_{0,x} - 1), \label{eq:C2} \\ \bar{j}_{1,x} = \bar{h}_1 (\bar{U}_{0,x} - 1) + \bar{h}_0 \bar{U}_{1,x}
\label{eq:C3}
\end{align}
%  -h_0 + \Ca^{-1} \frac{h_0^3}{3} \pderiv{(\Delta h_0)}{x}
% - \frac{1}{3} \left(3 h_1 + h^3_0(\Ub_0 \cdot \nabla) \Ub_0 - \Ca^{-1} h^2_0 \left( h_0\pderiv{(\Delta h_1)}{x} + 3 h_1 \pderiv{(\Delta h_0)}{x} \right) \right)
where the subscript $x$ denotes the component along the $x-$direction. The expressions for $\bar{h}_0$, $\bar{h}_1$, $\bar{\Ub}_0$ and $\bar{\Ub}_1$ are given in \eqref{ss0}, \eqref{ss1}, \eqref{eq:U0_final} and \eqref{U1_final} respectively.
We assume terms of $\mathcal{O}(\epsilon^2)$ or higher orders have negligible contribution to the flux. 
The net flux through a cross-section of the drop at an arbitrary location $x$ is 
\begin{equation}
\bar{J}_x = 2 \int_0^{\zeta_c \bar{x}} \bar{j}_x \;\; \mathrm{d}\bar{y}.
\label{eq:Jx}
\end{equation}
Substituting \eqref{eq:C2}, we obtain the net flux through any cross-section in the Stokes limit, in dimensionless form:
\begin{align}
\mathcal{O}(1):\quad \bar{J}_{0,x} = -2 \bar{x}^2 F_0(\zeta_c) \qquad \mathrm{where,}\quad F_0(\zeta_c) = \int_0^{\zeta_c} H (1 -H^2 f_x) \; \mathrm{d}\zeta.
\label{A}
\end{align}
The zero-net flux condition at this order yields a unique value of $H''(0)$ which solves for $H(\zeta)$ for a given opening angle. For example, when $H(0)=1.215$, the value of the function $F_0$ for various $H''(0)$ are shown in Fig \ref{fig:flux_H}. Only $H''(0)=-2.623$ satisfies the zero-net flux condition; the corresponding value of $\zeta_c = 1\; (\phi=45^\circ)$. Moreover, note that only for this particular choice of $H''(0)$ is the local flux across any point on the contact line also zero ($j_{0,x}(x,\zeta_c)=0$). The choice of $H(0)$ and $H''(0)$ obtained in this manner for a few values of $\phi$ are given in Table \ref{TableH}.

Similarly, the leading-order inertial flux is obtained by substituting \eqref{eq:C3} in \eqref{eq:Jx}
\begin{equation}
\mathcal{O}(\epsilon):\quad \bar{J}_{1,x} = -2 \bar{x}^3 F_1(\zeta_c), \quad \mathrm{where,} \quad  F_1(\zeta_c) =  \int_0^{\zeta_c} G (1 - H^2 f_x) - H^2g_x \; \mathrm{d}\zeta.
\label{A2}
\end{equation}
The flux at this order $\mathcal{O}(\epsilon)$ should also independently vanish in order to satisfy the macroscopic mass-flux condition. This condition should be evoked to determine the correct choice of the boundary conditions. Note that $\zeta_c$ is determined from the solution at $\mathcal{O}(1)$; In this example, we have $\zeta_c=1$ (see Fig. \ref{fig:flux_H}). Fig \ref{fig:flux_H}(d) shows the different values of the boundary condition $G''(0)$ for the case $\zeta_c=1$ and the correct value which satisfies the zero-flux condition is highlighted.  Moreover, note that the local flux across the contact line is also zero ($j_{1,x}(\zeta_c)=0$) only for this particular choice of $G''(0)$.
 
 %%%%%%%%%%%%%%%%%%%%%%%%%%%%%%%%%%%%%%%%%%%%%%%
%\clearpage
%\bibliography{Biblio.bib}
%%\bibliographystyle{unsrt}
%\bibliographystyle{unsrtnat}
%%%%%%%%%%%%%%%%%%%%%%%%%%%%%%%%%%%%

\end{document}